\documentclass[sigconf,table,xcdraw,screen]{acmart}
\usepackage[ruled,vlined]{algorithm2e}
\usepackage{graphicx}
\usepackage{paralist}
\usepackage{tabularx}
\usepackage{balance}
\usepackage{multirow}
\usepackage{multicol}
\usepackage{pbox}
\usepackage{mathtools}
\usepackage[TABBOTCAP]{subfigure}
\usepackage{algorithmic}
\usepackage{paralist}
\usepackage{tabularx}
\usepackage{url}
\usepackage{balance}
\usepackage{multirow}
\usepackage{multicol}
\usepackage{amsmath}
\usepackage{setspace}
\usepackage{verbatim}
\usepackage{float}
\usepackage[normalem]{ulem}
\usepackage{xspace}
\usepackage{amssymb}
\usepackage{ifthen}
\usepackage{pifont}
\usepackage{tikz}
\newcommand*\circled[1]{\tikz[baseline=(char.base)]{
            \node[shape=circle,draw,inner sep=0.5pt] (char) {#1};}}
\usepackage{hyperref}
\PassOptionsToPackage{bookmarks=false}{hyperref}

\newcommand{\CrashScope}{{\sc CrashScope}\xspace}
\newcommand{\CrashScopes}{{\sc CrashScope's}\xspace}
\newcommand{\GVT}{{\sc Gvt}\xspace}

\newcommand{\Gcat}{{\sc Gcat}\xspace}
\newcommand{\Gcats}{{\sc Gcat's}\xspace}

\include{macro}

\begin{document}

\title[Detecting and Summarizing GUI Changes in Evolving Mobile Apps]{Detecting and Summarizing GUI Changes\\ in Evolving Mobile Apps}

\author{Kevin Moran, Cody Watson, John Hoskins, George Purnell, and Denys Poshyvanyk}
\affiliation{%
  \institution{College of William \& Mary \\ Department of Computer Science}
  \streetaddress{P.O. Box 8795}
  \city{Williamsburg}
  \state{VA} 
  \country{USA}
  \postcode{23185}
}
\email{{kpmoran,cawatson,jbhoskin,gwpurn,denys}@cs.wm.edu}

\renewcommand{\shortauthors}{K. Moran, C. Watson, J. Hoskins, G. Purnell, and D. Poshyvanyk}

\begin{abstract}
	Mobile applications have become a popular software development domain in recent years due in part to a large user base, capable hardware, and accessible platforms.  However, mobile developers also face unique challenges, including pressure for frequent releases to keep pace with rapid platform evolution, hardware iteration, and user feedback.  Due to this rapid pace of evolution, developers need automated support for documenting the changes made to their apps in order to aid in program comprehension. One of the more challenging types of changes to document in mobile apps are those made to the graphical user interface (GUI) due to its abstract, pixel-based representation.  In this paper, we present a fully automated approach, called \Gcat, for detecting and summarizing GUI changes during the evolution of mobile apps.  \Gcat leverages computer vision techniques and natural language generation to accurately and concisely summarize changes made to the GUI of a mobile app between successive commits or releases.  We evaluate the performance of our approach in terms of its precision and recall in detecting GUI changes compared to developer specified changes, and investigate the utility of the generated change reports in a controlled user study.  Our results indicate that \Gcat is capable of accurately detecting and classifying GUI changes -- outperforming developers -- while providing useful documentation.
\vspace{-0.1cm}
\end{abstract}


\setcopyright{acmlicensed}
\acmPrice{15.00}
\acmDOI{10.1145/3238147.3238203}
\acmYear{2018}
\copyrightyear{2018}
\acmISBN{978-1-4503-5937-5/18/09}
\acmConference[ASE '18]{Proceedings of the 2018 33rd ACM/IEEE International Conference on Automated Software Engineering}{September 3--7, 2018}{Montpellier, France}
\acmBooktitle{Proceedings of the 2018 33rd ACM/IEEE International Conference on Automated Software Engineering (ASE '18), September 3--7, 2018, Montpellier, France}


\maketitle

\section{Introduction}
\label{sec:intro}

	Mobile application development has solidified itself as a prominent specialization for software engineers.  In fact, according to StackOverflow's 2018 survey of developers~\cite{so-survey}, over 20\% of respondents identified as mobile developers, making this the fourth most popular specialization overall, behind three different web development roles.  This popularity is sustained by several factors including a large and growing user base, performant hardware, powerful development platforms and APIs, and ease of software distribution through app marketplaces, just to name a few. 

	  Highly competitive app stores like Apple's App Store~\cite{apple-app-store} or Google Play~\cite{google-play} contain millions of apps, many of which implement similar functionality. In order to succeed in such marketplaces, developers need to ensure their application provides an engaging user experience and aesthetically pleasing user interface~\cite{design-importance}. Unfortunately, past studies have shown that designing and implementing effective GUIs can be a difficult task~\cite{Tucker:CSH04,Myers:CHD94,Nguyen:ASE'15}, especially for mobile apps~\cite{Moran:ICSE'18}. These difficulties are due in large part to challenges unique to the mobile development process that have been well documented in research literature~\cite{Joorabchi:ESEM'12} and include: (i) rapidly evolving platforms and APIs~\cite{Linares-Vasquez:FSE'13,Bavota:TSE15}, (ii) continuous pressure for new releases~\cite{Hu:EuroSys'14,Jones:2014}, (iii) inefficiencies in testing~\cite{Choudhary:ASE'15,Linares-Vasquez:ICSME'17a,Linares-Vasquez:ICSME'17}, (iv) overwhelming and noisy feedback from user reviews~\cite{Ciurumelea:SANER'17,DiSorbo:FSE'16,Palomba:ICSE'17,Palomba:ICSME'15}, and (v) market, device, and platform fragmentation~\cite{Han:WCRE'12,Wei:ASE'16,android-fragmentation}.

	Mobile GUIs are typically stipulated in files separate from the main logic of the app (\eg \texttt{\small .xml} for Android, and \texttt{\small .nib} or storyboards for iOS). These files delineate attributes of GUI components in relative terms (\eg display independent pixel \texttt{\small dpi} values) and are arranged according to a hierarchical structure (\ie a GUI hierarchy) to facilitate reactive design across fragmented device ecosystems. Reasoning about the actual rendering of a GUI using such an abstract definition in code is a difficult task. Conversely, collecting screenshots to discern visual changes is difficult, as it requires manual intervention and adept visual perception is needed to discern meaningful GUI changes.  Thus, it is clear that comprehending how GUI code affects the visual representation of an app requires mentally bridging a challenging abstraction gap. 

Furthermore, the design and implementation of a GUI for a mobile app is not a ``single cost'' task that is performed at the inception of development.  Instead, GUI-changes must evolve to keep pace with constant user feedback and the evolution of the prescribed design language and guidelines of the underlying mobile platform (\eg Android's transitions to differing versions of material design~\cite{material-design}), thus developers must constantly evolve an app's GUI to satisfy changing design requirements. This illustrates that there is a clear need for automated support in effectively \textit{documenting} GUI changes to help aid developers in time-consuming program comprehension tasks related to mobile app development. In particular, automated  summarization of \textit{visual} GUI-changes would allow for developers to more effectively comprehend the affect of code-based changes on the visual representation of a mobile GUI. 

	To assist developers in comprehending GUI changes in mobile apps, we introduce a fully automated approach aimed at detecting, classifying, and summarizing visual GUI changes between subsequent app versions. Our approach, called \Gcat (\textbf{G}UI \textbf{C}hange \textbf{A}nalysis \textbf{T}ool), is triggered upon a specified commit to a mobile app's version control system and performs a GUI differentiation analysis. This process begins by automatically executing the target app, extracting a representative set of screenshots and GUI-metadata, and comparing these to similar files extracted from a previous version of the same app using computer vision techniques. \Gcat then generates a comprehensive report describing GUI changes that includes annotated screenshots, a natural language summary of GUI changes, and a visualization of matching segments of each screen's GUI hierarchy. 

	We performed an extensive evaluation of \Gcat across several different quality attributes. First, we empirically examined the performance of \Gcat in terms of (i) automatically extracting\slash filtering\slash matching screens and (ii) detecting and classifying GUI changes from a set of 31 mobile apps from the F-Droid~\cite{fdroid} repository of open source apps. Next we performed a user study measuring developers' performance in detecting and classifying mobile app GUI changes, and the perceived usefulness of the GUI change summarization reports produced by \Gcat. Our results indicate that \Gcat is able to (i) accurately and automatically extract, filter and match screens between subsequent versions of Android apps, (ii) effectively detect and summarize GUI-changes, (iii) outperform developers in terms of identifying, detecting, and classifying GUI changes, and (iv) automatically generate GUI summarization reports that developers found useful in comprehending GUI changes. In summary, this paper makes the following contributions:

\begin{itemize}
	\item{ We introduce \Gcat, a fully automated approach for detecting, classifying, and summarizing GUI changes in evolving mobile apps;}
	\item{We conduct a comprehensive evaluation of \Gcat that measures its detection and classification performance compared to developers, and the perceived usefulness of \Gcat reports;}
	\item{We derive a sizable dataset of GUI changes isolated from real FOSS apps which can facilitate future research in program comprehension related to mobile GUIs;}
	\item{We make available an online appendix~\cite{online-appendix} that includes additional materials such as examples of reports generated by \Gcat, an open source version of our approach, and all study data to facilitate reproducibility.}
\end{itemize}
\vspace{0.3cm}


\vspace{-0.5cm}
\section{Background \& Problem Statement}
\label{sec:background}

\begin{figure}
\centering
\includegraphics[width=0.9\columnwidth]{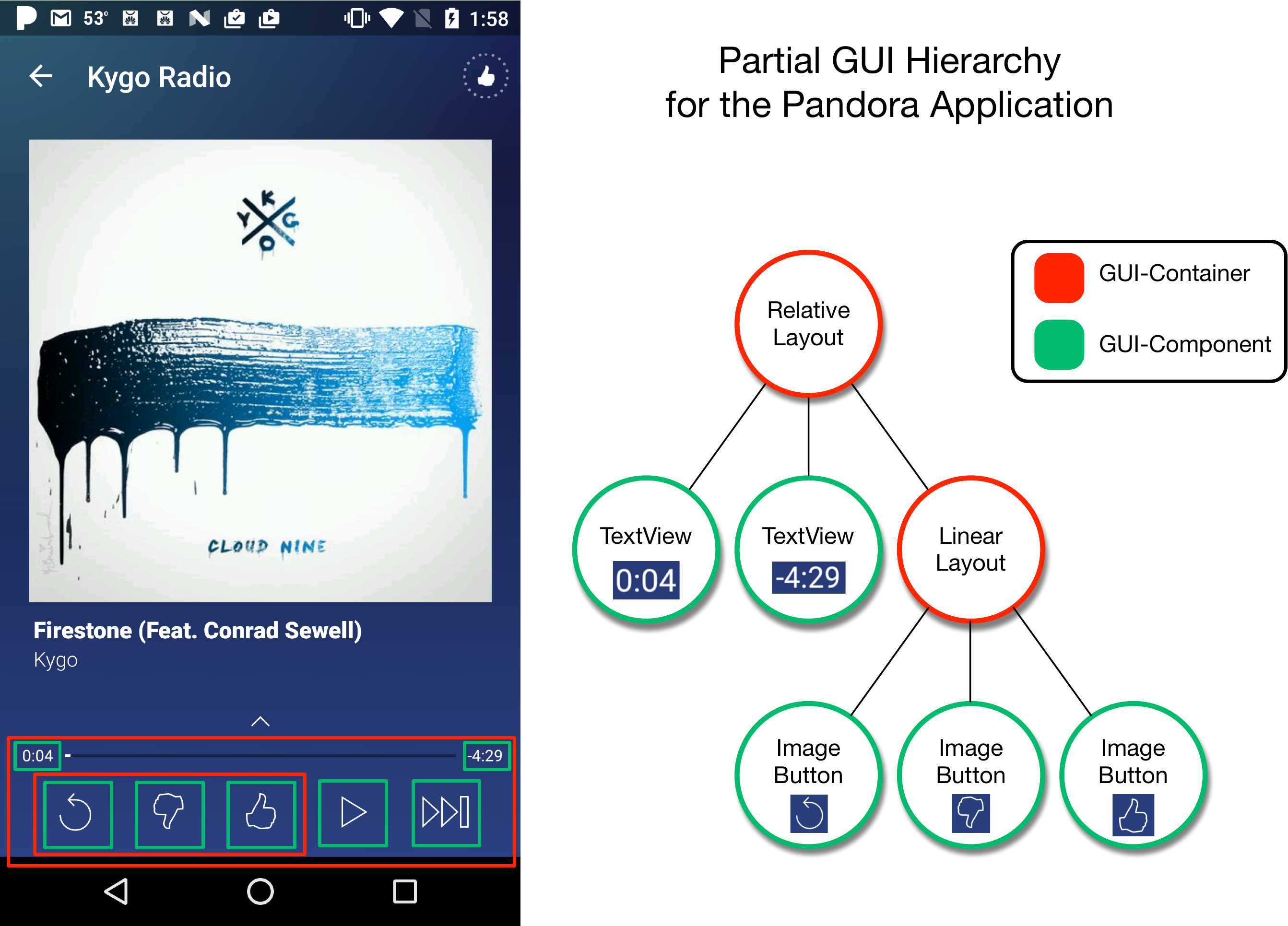}
\vspace{-0.1cm}
\centering
\caption{Illustration of the GUI Structure of the Pandora Android Application}
\label{fig:gui-structure}
\vspace{0.1cm}
\end{figure}

	In general, the goal of the approach set forth in this paper is to automatically detect, classify, and summarize changes that occur in the GUI of an evolving mobile app.  Our approach is currently implemented for Android (the most widely used OS in the world~\cite{statista-mobile-market-share-q2-2017}) despite being applicable to other platforms such as iOS.  Thus, in this paper we examine the  principles of mobile GUIs and GUI changes in the context of Android.  

\subsection{Mobile GUI Fundamentals}
\label{subsubsec:mobile-guis}

	In the context of Android there are two basic logical constructs that comprise the GUI of a mobile app, which are illustrated in Figure~\ref{fig:gui-structure}.  The first of these is a \textit{GUI-component}. GUI-components (used interchangeably with the term ``component" in this paper) have been defined in prior work as ``\textit{atomic graphical elements with pre-defined functionality, displayed within the GUI of a software application}"~\cite{Moran:ArX'18}. In the context of Android there are several different types of components, such as \texttt{\small TextViews}, \texttt{\small Buttons}, and \texttt{\small NumberPickers}.  Each of these serves a distinct set of purposes. For instance, a \texttt{\small Button} is typically used to trigger certain functionality from the code, and a \texttt{\small NumberPicker} allows a user to select from a pre-defined range of numbers as input.  In addition to their \textit{type} there are also several stylistic details that define a component, such as a displayed image, colors, or font.  Two \texttt{\small TextView} components and three \texttt{\small ImageButton} components are shown highlighted in green as part of the GUI for the popular Pandora Music app in Figure \ref{fig:gui-structure}.  As this figure shows, each component has a bounding box that stipulates the area occupied by the component, this is typically defined by spatial coordinates such as the $x$ and $y$ coordinates of the top left-hand corner of the box, and its $width$ and $height$.

\begin{table*}[]
\centering
\caption{The taxonomy of GUI changes used in the development and evaluation of \Gcat}
\vspace{-0.2cm}
\fontsize{7.5pt}{10}
\selectfont
\label{tab:gui-change-taxonomy}
\begin{tabular}{|l|l|l|}
\hline
\textbf{Change Category}         & \textbf{Specific Change}           & \textbf{Description}                                                                                                      \\ \hline
\multirow{3}{*}{Text Change}     & Text Change                        & The text content of a component from a previous version of the app does not match a later version                         \\ \cline{2-3} 
                                 & Font Change                        & The text font of a component from a previous version of the app does not match a later version                            \\ \cline{2-3} 
                                 & Font Color Change                  & The text font color of a component from a previous version of the app does not match a later version                      \\ \hline
\multirow{6}{*}{Layout Change}   & Vertical Translation               & The location of a component was translated in the vertical direction between  versions of an app                \\ \cline{2-3} 
                                 & Horizontal Translation             & The location of a component was translated in the horizontal direction between  versions of an app              \\ \cline{2-3} 
                                 & Vertical Size Change               & The size of a component was changed in the vertical direction between  versions of an app                       \\ \cline{2-3} 
                                 & Horizontal Size Change             & The size of a component was changed in the horizontal direction between  versions of an app                     \\ \hline
\multirow{6}{*}{Resource Change} & Image Color Change                 & The color of an image associated with a component changed between  versions of an app                           \\ \cline{2-3} 
                                 & Removed Component                  & A component was removed between  versions of an app                                                             \\ \cline{2-3} 
                                 & Added Component                    & A component was added between  versions of an app                                                               \\ \cline{2-3} 
                                 & Image Change                       & The image associated with a component was changed between  versions of an app                                   \\ \cline{2-3} 
                                 & Component Type Change              & The type of a component changed between  versions of an app                                                                    \\ \hline
\end{tabular}
\vspace{-0.3cm}
\end{table*}

	However, GUI-components are not the only building block that comprise a mobile GUI.  There also exist \textit{GUI-containers}, which have been succinctly defined in prior work as ``\textit{A logical construct that groups members of GUI-components and typically defines spatial display properties of its members}"~\cite{Moran:ArX'18}.  Thus, GUI-containers are largely meant to help provide a spatial structure to the GUI and define stylistic details regarding the background or canvas upon which GUI-components are rendered. GUI-components are typically rendered on a screen according to the spatial properties of their containers, rather than predefined screen coordinate values.  This allows for a more flexible design that can fluidly adapt between devices with different display dimensions and densities.  Two GUI-containers, a \texttt{\small RelativeLayout} and a \texttt{\small LinearLayout} are highlighted in red for the Pandora App in Figure \ref{fig:gui-structure}.

	When taken together, GUI-components, and GUI-containers compose a \textit{GUI-hierarchy}, which typically takes the form of a rooted tree where smaller components and containers exist within a single container that serves as the root of the hierarchy.  In Figure \ref{fig:gui-structure}, a partial GUI-hierarchy for the Pandora app is illustrated as a tree.  In this hierarchy, the \texttt{\small RelativeLayout} serves as the root node with other GUI-components and containers filling out the tree. As stated earlier, the GUI-hierarchies for mobile apps are typically defined in a domain specific language outside of the functional code of an app.  In Android, properties of the GUI are stipulated in xml files in the app resource directory (\eg \texttt{\small /res/layout}) using a domain specific xml format.  When an Android app's GUI is rendered on a device screen, metadata describing the GUI (including information such as the coordinates of rendered components, their types, and whether or not they are interactive) can be read from a device using the \texttt{\small uiautomator} framework.  It is important to note that there are distinct differences between the static and dynamic representations of an app's GUI. Full information regarding the appearance of a GUI cannot be gleaned from the static-code representation alone, as this information is defined in relative terms and the GUI must be interpreted and instantiated for target screen attributes.  Furthermore, components such as lists can be dynamically populated at runtime, which impacts GUI appearance.  

\vspace{-0.3cm}
\subsection{Evolutionary GUI Changes}
\label{subsubsec:gui-changes}

\begin{figure}
\centering
\vspace{0.1cm}
\includegraphics[width=\columnwidth]{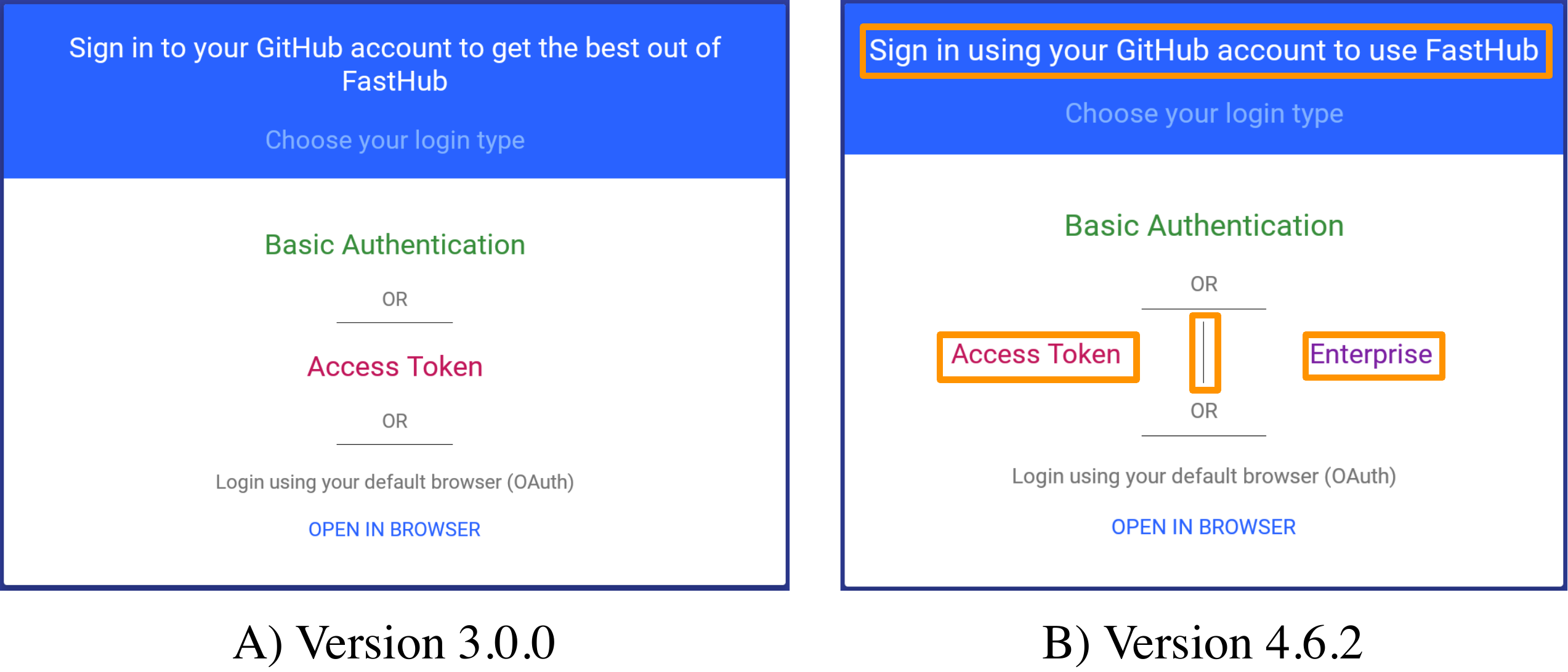}
\vspace{-0.7cm}
\centering
\caption{Illustration of GUI changes in FastHub}
\label{fig:gui-changes}
\end{figure}

	Now that we have described the basic building blocks of mobile GUIs, it is important to understand how GUI-changes affect these building blocks and how they might be logically categorized.  At a high-level, a \textit{GUI-change} can be described as any modification to the spatial or stylistic properties of a GUI-component or container.  There are a finite number of logical manners in which components can be altered between app versions. In order to accurately describe GUI changes, it is important to stipulate different categorizations of changes that might occur.  

	To do this, we look to past work on detecting design violations in mobile apps~\cite{Moran:ICSE'18}. A \textit{design violation} in the context of mobile apps has been defined as a mismatch between the attribute vectors of two GUI-components that exist both in a mobile GUI mock-up and implementation, where the attribute vectors can be represented as a a four-tuple in the form \textit{($<$x-position,y-position$>$, $<$height,width$>$, $<$text$>$, $<$image$>$)}~\cite{Moran:ICSE'18}. In this work the authors performed a grounded-theory survey on an industrial dataset of design violations and derived a taxonomy.  Given that in this work, a design violation essentially describes a \textit{change} in a mobile GUI (albeit one introduced erroneously by a developer), we adapt this taxonomy to describe \textit{GUI-changes} that surface between subsequent versions of a mobile app.

\begin{figure*}
\centering
\vspace{-0.2cm}
\includegraphics[width=\linewidth]{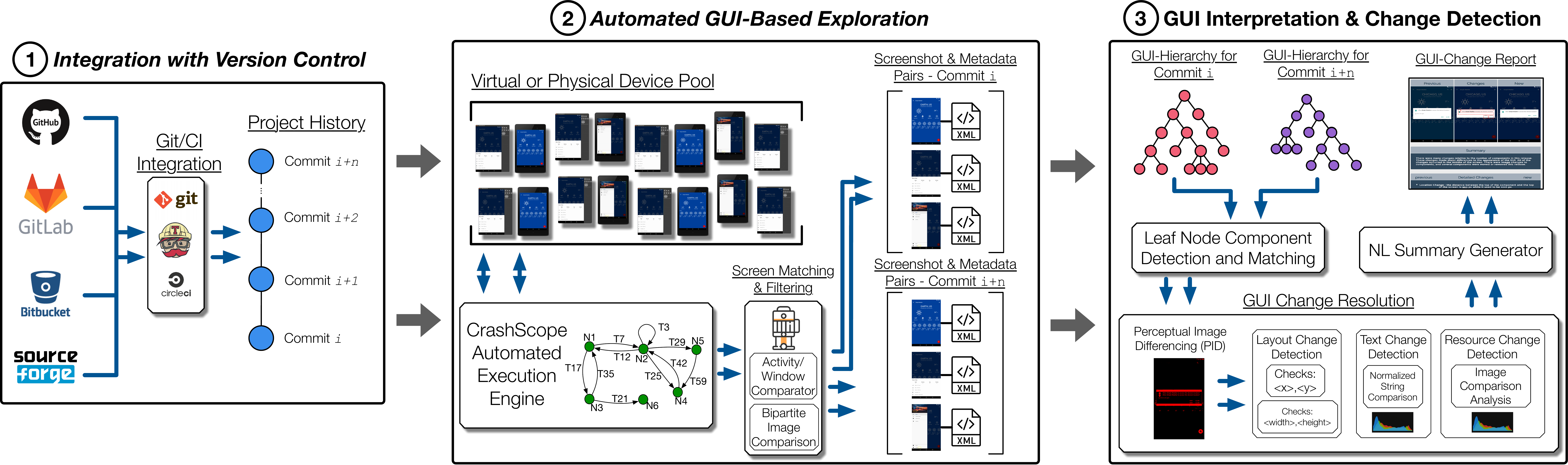}
\vspace{-0.6cm}
\centering
\caption{Overview of the \Gcat Approach}
\label{fig:approach}
\vspace{-0.4cm}
\end{figure*}

	Our GUI-change taxonomy is described in Table \ref{tab:gui-change-taxonomy} and consists of three main categories: (i) \textit{Text Changes} that concern differences in text displayed by components, (ii) \textit{Layout Changes} that concern differences between the spatial properties of components, and (iii) \textit{Resource Changes} that describe phenomena such as missing or added components, or differences between utilized images or colors.  Each of these three main categories has a subset of specific change categories, which directly describe a GUI-change. It should be noted that more than one GUI-change can apply to a single GUI component.  For instance, a component might change in size and location between app versions.  Some examples of GUI-changes between subsequent app versions of the popular FastHub GitHub client are illustrated in Figure \ref{fig:gui-changes}.  For example, the ``Access Token" \texttt{\small TextView} component exhibits a \textit{Layout Change}, whereas the the ``Enterprise" \texttt{\small TextView} component represents an \textit{Added Component} change.  The \texttt{\small TextView} component, which displays ``Sign in using your GitHub account to use FastHub", exhibits two change types, namely a \textit{Text Change} and a \textit{Size Change}.  

	Given this background on mobile GUIs and our GUI-change taxonomy our problem statement can be formulated as follows: 

\noindent \textbf{\textit{Problem Statement:}} \textit{Given an Android app with a change history $V_1 , V_2 ... V_i$, our approach aims to automatically extract screenshots and GUI metadata for two versions $V_i$ and $V_k$ such that $k >i$. Then our approach aims to filter corresponding screens between the two versions and detect, classify, and summarize GUI changes between corresponding pairs of screens.}

\vspace{-0.3cm}
\section{The \Gcat Approach}
\label{sec:approach}

	There are three main components of the \Gcat approach depicted in Figure \ref{fig:approach}, (i) \textit{Version Control Integration}, (ii) \textit{Automated GUI-Based Exploration}, and (iii) \textit{GUI Interpretation and Change Detection}.  \Gcat is able to analyze subsequent commits from a software repository that utilizes a version control system such as Git, and automatically compiles and executes target commits. It then filters and matches screens discovered during automated exploration of a target app's GUI, and finally detects and reports changes related to GUI-components.  \Gcat was implemented for the Android platform and is written in Java.  It was designed to allow for incorporation into Continuous Integration and version control systems to facilitate fully automated generation of documentation. From a developer's perspective, \Gcat would simply need to be installed on a developer's machine or integrated into continuous integration (CI) pipeline, and frequency of analysis specified (\eg running on each commit, or major releases, etc.). Then developers could view the web-based GUI-change reports either locally, or via a CI system, in order to more effectively comprehend the GUI-changes between subsequent app versions.  In this section we describe each component of the \Gcat approach.

\vspace{-0.3cm}
\subsection{Integration with Version Control}
\label{subsec:version-control}

	In order to provide practical automated documentation of GUI changes as a mobile app evolves, \Gcat can take advantage of the version control system of a target mobile app's software repository. Furthermore, \Gcat could be triggered in a Continuous Integration Pipeline such as TravisCI~\cite{travis-ci} or GitLab's CI framework~\cite{gitlab-ci}, as illustrated in Figure \ref{fig:approach}-\circled{1}.  In order to derive and document changes in a change report, \Gcat accepts two subsequent commits $i$ and $i+n$ where $n$ represents the number of commits between analyzed versions. When a new repository is created, or \Gcat is added to the CI system of an existing repository, \Gcat automatically compiles and performs automated GUI-based exploration of the most recent commit of the project and caches extracted screenshots and GUI-related metadata. Additionally, a developer may stipulate that \Gcat analyze subsequent pairs of historic commits. The choice of how frequently to run \Gcat is left to the developer. In Section \ref{sec:study-design} we detail our experimental methodology for deriving subsequent commits.

\vspace{-0.3cm}
\subsection{Automated GUI-Based Exploration}
\label{subsec:gui-explore}

	Once two commits have been isolated from an Android app's repository, screenshots, and metadata describing the programmatic structure of the GUI-hierarchy associated with these screenshots must be automatically extracted. Then, once a set of screens has been extracted, \textit{corresponding} screens from the pair of target commits must be matched with one another, and redundant screens must be filtered out, in order to reduce the information burden on developers. This process is illustrated in Figure \ref{fig:approach}-\circled{2}.

\subsubsection{Automated GUI-Exploration}

	In order to automatically explore the GUI of a target app, \Gcat makes use of the \CrashScope~\cite{Moran:ICST'16,Moran:ICSE-C'17} GUI-exploration engine. \CrashScopes automated exploration simulates touch events on a mobile device or emulator to explore the screens of a target app.  To do this effectively, the \CrashScope engine performs a systematic, depth-first search (DFS) exploration of an app's GUI that has been shown to achieve comparable coverage to other testing approaches \cite{Moran:ICST'16}. During this exploration process, the GUI of an app is analyzed in real time using Android's \texttt{\small uiautomator}~\cite{uiautomator} framework. Interactive components are identified, and an event-flow model of an app is constructed in an online manner. DFS exploration proceeds according to a given set of parameters known as an \textit{exploration strategy}. In our adaptation of \CrashScopes exploration engine for \Gcat, we utilized two variations of the DFS GUI traversal, a \textit{top-down} variation where interactive components are exercised from the top of the screen down, and a \textit{bottom-up} variation, where interactive components are exercised from the bottom of the screen up.  Two variations of text-input strategies were utilized, one strategy generated \textit{expected} text by inputting allowable characters according to parameters of a given text field, and another strategy generated \textit{no text} to be input to text fields. We chose not to implement other strategies from the original \CrashScope execution engine due to the fact that these strategies were more likely to discover crashes from a target app, and our objective in \Gcat is not crash detection, but rather state exploration. The selected exploration and text input strategies exhibited higher coverage in past work~\cite{Moran:ICST'16}. For each action that \CrashScope executes on a device, a screenshot, and dump of the GUI metadata from  \texttt{\small uiautomator} are saved before and after the action's execution. This set of screenshots and GUI-metadata are then passed to the screen matching and filtering procedure. 

	Currently, \Gcat only supports GUI comparisons between corresponding screens captured on the same device.  However, it should be noted that the automated exploration for \Gcat can be run on a concurrent set of virtual Android devices that simulate a range of screen sizes/densities in order to extract GUI information for a predefined set of device configurations. \Gcat reports can then be generated for \textit{corresponding} screen pairs on a per-device basis. Furthermore, \Gcat could be adapted to utilize a set of pre-specified automated GUI tests using a test wrapper that captures screenshots and \texttt{\small uiautoamtor} files after each test case step.

\subsubsection{Screen Matching and Filtering}

Using the screenshot and metadata pairs that can be generated for a given pair of commits, \textit{corresponding screens} (screens that retain highly similar intended functionality) can be identified between commits for which useful change reports can be derived. We model this process as a bipartite matching problem, where the cost of an assignment $C$ between any two screen pairs $s_1, s_2$ is the sum of two values: 
\begin{equation}
\label{eq:screen-fitlering}
C(s_1, s_2) = CD + BBOX_{diff}
\end{equation}
where $CD$ is equal to the Euclidean color distance between the two images, and $BBOX_{diff}$ is equal to the normalized pixel difference between two binary images $b_1$ and $b_2$, created by drawing white filled rectangles corresponding to the bounding boxes of the leaf node components onto a black silhouette of the screen. Each bounding box will only be drawn if its total area is less than 100k pixels, to avoid large overlay components from affecting the analysis.

Both constituents of the assignment score are orthogonally beneficial: $CD$ is able to capture pure visual similarity, but is a poor measure of matching potential in examples where there are a large number of color changes. For these cases, we need a way to utilize the structural information of the screenshots provided by $BBOX_{diff}$. Both sets of nodes in our bipartite graph correspond to the screen-xml pairs for their respective commits. The edge weights between each node are equal to $C(s_i, s_j)$ for all $i$, $j$ in each set. Once the graph is constructed as an adjacency matrix $M$, we find a matching $\alpha$ such that it minimizes the sum cost of all assignments.

The optimization algorithm used in our implementation runs in $O(n^3)$ time. In addition, the sets of screens from target pairs of subsequent commits may be quite large. Thus, in order to make this process dramatically more efficient, we defined a lightweight heuristic to cut back on superfluous screens and reduce the size of the sets. During each step in the automatic GUI exploration, the name of the current activity is recorded, as well as the name and type of the currently active window (\eg \texttt{\small FRAGMENT}, \texttt{\small POPUP}). This information was extracted at each step of the execution using the \texttt{\small adb shell dumpsys window windows} command. Using this information, we filter our screen sets such that only screen-xml pairs that represent the first occurrence during the automated execution of a unique (activity, window) pair are kept. All others are discarded. From a developer's perspective, GUI-change reports will only be generated for matched screens, however, \Gcat could also be configured to allow a developer to examine unmatched screens pairs and trigger the change analysis for these pairs. 

\vspace{-0.2cm}
\subsection{GUI Interpretation \& Change Detection}
\label{subsec:gui-change-detection}

	Once corresponding screen pairs between a target pair of commits have been extracted using \Gcats automated GUI exploration and screen matching and filtering techniques, \Gcat then needs to identify the GUI-changes that occurred between these screen pairs.  To do this, \Gcat decodes the hierarchical representation of the GUI represented in a given screenshot using data from \texttt{\small uiautomator} xml files. It then identifies and matches corresponding GUI-components between screen pairs, analyzes corresponding components for changes, and classifies these changes. Finally, an html-based GUI change report is generated complete with images and natural language descriptions of changes. This process is visualized in Figure \ref{fig:approach}-\circled{3}.

\subsubsection{GUI Hierarchy Construction \& Component Matching}

	For a given corresponding screen pair \Gcat parses the \texttt{\small uiautomator} xml files associated with each screenshot and constructs a tree-based representation of the GUI-hierarchy. It then parses and stores collections of \textit{leaf node} components for each screen, including several attributes such as location information (\eg \texttt{\small <x,y><width,height>}) and the component type (\eg \texttt{\small ImageButton}). As stated earlier, \Gcat reports GUI-changes according to \textit{leaf node} components, as they tend to also reflect changes to container components. Thus, \Gcat employs a \textit{k}-nearest neighbors matching procedure based on spatial component information that has proven successful in past work on reporting GUI design violations for mobile apps~\cite{Moran:ICSE'18}. During this procedure, each component is matched against its closest neighbor according to the following simialrity score: 
\begin{equation}
\label{eq:MatchingFunction}
\gamma = (|x_m - x_r| + |y_m - y_r| + |w_m - w_r| + |h_m - h_r|)
\end{equation}
\noindent where a smaller $\gamma$ represents closer matches. The $x,y,w$ and $h$ variables correspond to the $x$ \& $y$ location of the top and left-hand borders of the bounding rectangle, and the height and width of the bounding rectangles for components respectively.

\subsubsection{GUI-Change Resolution}

	After corresponding pairs of leaf node components have been identified, \Gcat must then detect GUI changes between screens. \Gcat first employs Perceptual Image Differencing (PID), an image differencing algorithm modeled after the human visual system that has been successfully applied in past research on detecting GUI differences \cite{Moran:ICSE'18,Mahajan:ICST'15,Mahajan:ICST'16}. PID helps to identify a set of potential changes based on visual differences between images. Then, each of these potential GUI changes is analyzed further to determine the specific type of change to report. This in-depth analysis varies depending upon the type of change. These analyses have been adapted from prior work on detecting GUI design violations~\cite{Moran:ICSE'18} to work with GUI metadata from corresponding screens extracted from commits of a target app.

\noindent\textbf{\textit{Layout Changes}}: Identifying \textit{Layout Changes} is relatively straightforward. \Gcat simply compares the \texttt{\small <x,y>} and \texttt{\small <width,hieght>} values for each pair of corresponding leaf components. If differences in $x,y,width$ or $hieght$ vary by more than a threshold $LC$, then a Layout Change is reported.

\noindent\textbf{\textit{Text Changes}}: There are three different types of text changes: (i) Font Color change, (ii) Font Style change, and (iii) Text Content change. Each of these specific types is detected in a different manner, but all utilize cropped images for each pair of potentially changed text components by cropping an image from both previous and subsequent screenshots according to the bounding boxes of the components in question.  To check for a Font Color change,  a color histogram (CH) is constructed for each cropped image by accumulating instances of all unique RGB pixel values.  \Gcat then calculates the normalized Euclidean distance between these color histograms, and if the distance is greater than a threshold $FC$ a Font Color change is reported.  If the color histograms do match, then a Font Style change is reported.  To detect changes in Text Content, strings between text components are pre-processed to lowercase, spaces are removed, and the resulting strings are compared. If the string values do not match, a change is reported. Our implementation of \Gcat uses an $FC$ value of $85\%$.

\noindent\textbf{\textit{Resource Changes}}: \Gcat is able to report 5 different types of Resource changes including: (i) Added Components, (ii) Removed Components, (iii) Image Color changes, (iv) Image Changes, and (v) Component Type changes. Leaf node components that are added to a subsequent version of an app correspond to components without a matched corresponding component. Thus, these are reported as Added Component changes. Likewise, Missing Components are those components from the previous version of the app that were not able to be matched to components in the subsequent version. Image Changes are detected by extracting cropped images of components in question from screenshots of both versions of an app. Then, these cropped images are converted to a binary color space (\eg black and white) and PID is run again. If the images do match according to PID within a threshold $IC$ then an image change is reported. Otherwise a color change is reported. In our implementation of \Gcat $IC=20\%$.

\vspace{-0.1cm}
\subsubsection{Natural Language Summary Generation}  

	The GUI change reports generated by \Gcat contain a NL summary, as well NL descriptions of each identified change. \Gcats natural language summaries of \textit{all} GUI-changes include a description of both \textit{what} happened, as well as \textit{where} it happened. To do this, \Gcat identifies the parts of a given screen that contain the most changes. First, the screen is divided into a congruent 3x3 grid and changes are assigned to each grid region. If no grid section in the 3x3 division contains a majority of changes, the screen is divided into a congruent 2x2 gird and the process is repeated. This helps to inform the NL description of \textit{where} changes occurred.

	After changes are isolated to particular areas of the screen, they need to be effectively summarized. We use a heuristic-based approach for general summarization. Each change is described by three characteristics: 1) \textit{Level} - a string describing how much the GUIs changed visually; 2) \textit{Location} -  the location on the screen that changed the most; and 3) \textit{Amount} - a string describing the number of changes made to the GUI.

\begin{figure}
\centering
\includegraphics[width=0.85\columnwidth]{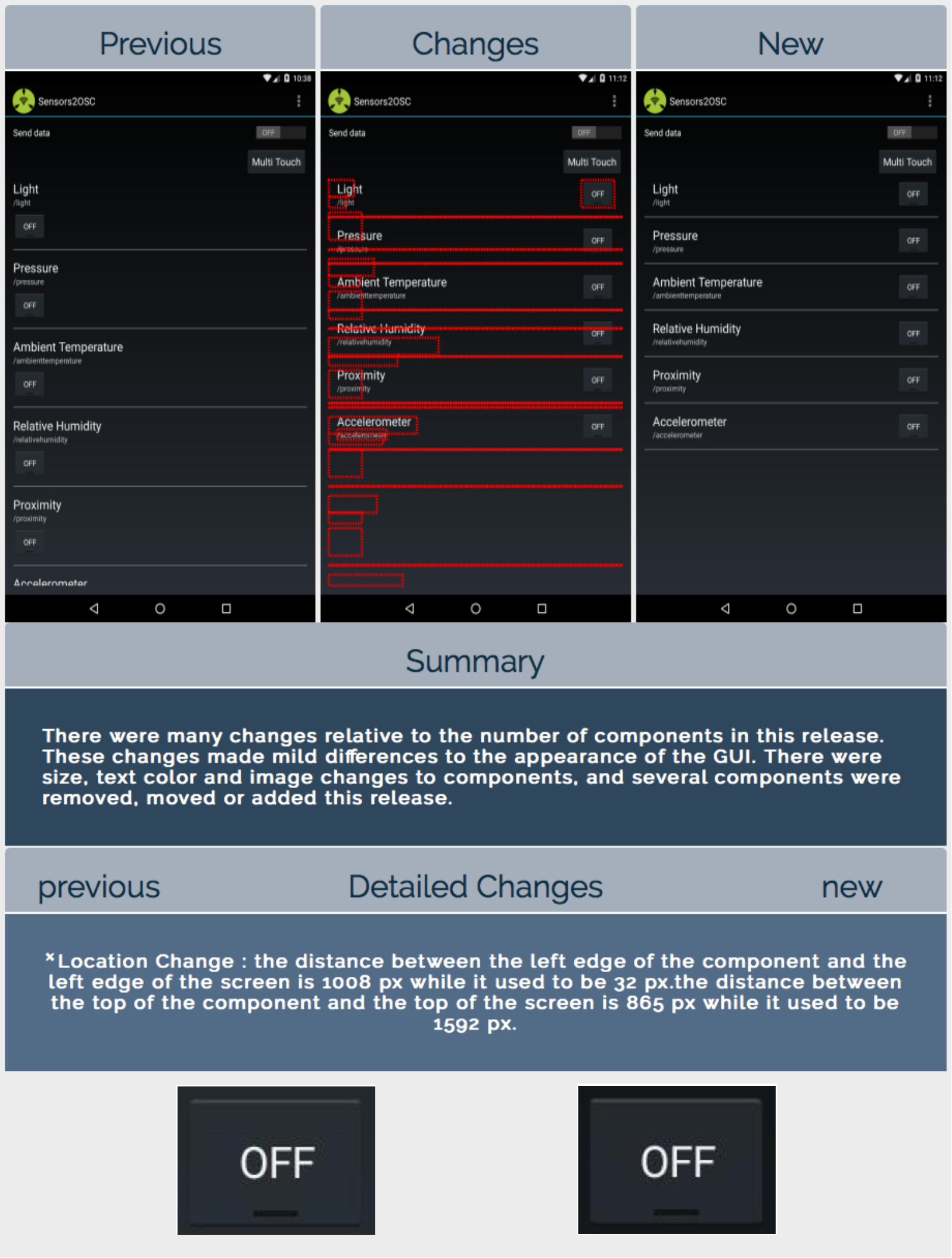}
\vspace{-0.3cm}
\centering
\caption{Partial Example of a Report Generated by \Gcat}
\label{fig:example-report}
\end{figure}

Depending on the values of the aforementioned characteristics our process determines which form the template will take. An example summary is given in Figure~\ref{fig:example-report}. We forgo an enumeration of the template and potential combinations due to space limitations, however, this is shown in our appendix. NL descriptions of individual GUI-change types are generated according to different templates specific for each change type.  

\vspace{-0.1cm}
\subsubsection{Report Generation}

	\Gcat generates \texttt{\small html} based reports that enumerate GUI changes in four major ways, three of which are illustrated by the example report in Figure \ref{fig:example-report}. The first of these is a set of full screenshots depicted at the top of the report, where the previous and subsequent screens are shown on the left and the right respectively, and the middle screenshot highlights changes from the perspective of the previous version screenshot. The second piece of information reported is the NL summary of changes in the GUI. The third piece of information is a list of detailed changes on a component-by-component basis. These include both a NL description and, if clicked on, a side by side comparison of the components in the old and new version of the app. Finally, the fourth piece of information (not shown in Figure \ref{fig:example-report}) is the maximum common spanning tree of the screen pair GUI hierarchies.

\vspace{-0.2cm}
\section{Design of the Experiments}
\label{sec:study-design}


	The overarching goal of \Gcat is to detect, classify and summarize GUI changes that occur in mobile apps as they evolve. Thus, to evaluate \Gcat, we carry out an empirical study aimed at investigating the \textit{performance} of the approach, and a user study aimed at analyzing \Gcats \textit{usefulness} to developers. To this end we explore the following four RQs:

\begin{itemize}
	\item \textbf{RQ$_1$}: \textit{How well does \Gcats screen matching and filtering procedure function?} 
	\item \textbf{RQ$_2$}: \textit{How well does \Gcat perform in terms of detecting and classifying GUI changes that occur during the evolution of mobile apps?} 
	\item \textbf{RQ$_3$}: \textit{Is \Gcat able to more accurately detect and classify GUI changes in evolving mobile apps compared to manual efforts from developers?}
	\item \textbf{RQ$_4$}: \textit{Do developers find \Gcat reports useful for documenting and summarizing GUI changes in evolving mobile apps?}
\end{itemize}

	In the context of our study, RQ$_1$, RQ$_2$, and RQ$_3$ are directed toward quantitatively measuring how well \Gcat performs in terms of extracting screens and detecting and classifying different types of GUI changes that occur during the evolution of Android apps.  RQ$_4$ is aimed at qualitatively measuring the perceived usefulness of \Gcat reports by collecting feedback regarding the user experience and preferences. To collect user data to help answer RQ$_3$ \& RQ$_4$ we conducted a user study in the form of an online survey.

\subsection{Study Context}
\label{subsec:study-context}

	In order to evaluate \Gcat, we required a set of popular subject applications from which a collection of GUI changes for particular screens between subsequent app versions exist.  To derive this set of screens, we utilized a set of 31 applications from FDroid~\cite{fdroid}. 

	To collect these apps, three authors manually crawled orthogonal sections of FDroid and downloaded the set of available release \texttt{\small apks} for each app.  In order to facilitate controlled experimentation and ensure a sizable set of screen pairs with existing GUI changes, the same authors launched subsequent versions of the apps on concurrent Nexus 7 (2013) emulators running Android 6.0 from Genymotion \cite{genymotion}, and ensured that at least one corresponding screen pair between the two versions exhibited a GUI change. Apps without any version pairs that could be launched on the emulator, that were hybrid apps, used non-standard components, or that did not exhibit any GUI changes were discarded.  This process resulted in a set of 62 \texttt{\small apks} corresponding to program versions from 31 apps. We provide detailed information about these apps, and make all of our study data available in our online appendix~\cite{online-appendix}.

\vspace{-0.2cm}
\subsection{RQ$_1$: Evaluating \Gcats Screen Matching and Filtering}
\label{subsec:rq1-methodology}
 
To measure how well \Gcats screen filtering and matching procedure function, we ran each of the 62 \texttt{\small apks} extracted for the study through the systematic automated input generation approach derived from the \CrashScope. The average time per app for running this exploration strategy and extracting the screenshots and GUI metadata is 39.46 minutes per app. However, it should be noted this process is completely automated and can be run passively in the background. We then measured two metrics: (i) the percentage of filtered screens ($FS$), and (ii) the matching precision ($MP$).  The $FS$ metric measures the number of redundant screens filtered and the $MP$ metric illustrates the number of correctly matched corresponding screens. More formally, these metrics can be represented as: 
\vspace{-0.1cm}
\begin{equation}
\label{eq:precision}
FS = \frac{TS -FS }{TS}\times 100 \quad MP = \frac{T_p}{T_p+F_p}
\end{equation}

\noindent where $TS$ is the total number of screens discovered by \CrashScope, $FS$ is the number of screens filtered by \Gcat, $T_p$ is the number of correctly matched screens, and $F_p$ is the number of incorrectly matched screens. One author examined the matched screens pairs from \Gcat in order to determine the $T_p$ and $F_p$, whereas the other metrics can be calculated automatically.

\vspace{-0.2cm}
\subsection{RQ$_2$: Measuring the Performance of \Gcat}
\label{subsec:rq2-methodology}

	The main \textit{goal} of this RQ is to examine how well \Gcat performs in terms of detecting and classifying real-world mobile GUI changes. In a practical use case of \Gcat, the entire GUI-change report generation process is automated, from the extraction of corresponding screen pairs, to the report generation.  Thus, in investigating this RQ we aimed to emulate this automated context by using the output of \Gcats screen matching and filtering procedure carried out as part of the previous RQ$_1$.

	\Gcats screen filtering/matching procedure resulted in a set of screen pairs consisting of \texttt{\small$<$screenshot,GUI-metadata$>$} tuples for corresponding screens between differing application versions. \Gcat was then applied to screen pair tuples that were correctly matched and the GUI-change summarization reports were generated. During the generation process, we also measured the time taken by \Gcat to generate each report.

	To measure the performance of \Gcat in detecting and classifying GUI-changes, three metrics were calculated: (i) the Detection Precision ($DP$), (ii) the Classification Precision ($CP$), and (iii) the Recall ($R$).  The $DP$ measures how well \Gcat can detect GUI changes, whereas the $CP$ measures how well detected changes are classified into their corresponding types. We make this distinction because \Gcat is capable of detecting, but incorrectly classifying component changes into their proper types. $DP$, $CP$ and $R$ were measured as:
\vspace{-0.1cm}
\begin{equation}
\label{eq:precision}
DP,CP = \frac{T_p}{T_p+F_p}	\quad R = \frac{T_p}{T_p+F_n}
\end{equation}

\noindent where for $DP$, $T_p$ represents GUI changes that were detected by \Gcat, and for $CP$, $T_p$ represents GUI changes that were both detected and correctly classified in their proper type.  For each of these metrics, $F_p$ corresponds to detected GUI-changes that either did not exist or that were misclassified respectively.  For Recall, $T_p$ represents GUI changes that were correctly detected and $F_n$ represents existing GUI changes in the ground truth that were not detected by \Gcat. Due to the cost of calculating these metrics, explained below, we randomly sampled 18 screen pairs from the correctly matched corresponding screens to answer RQ$_2$. To facilitate this, we ran each of the screen pairs through PID and ranked them in three \textit{GUI-change groups} ({\sc{High}}, {\sc{Medium}}, and {\sc{Low}}) according to the rank of the percentage of difference pixels reported by the PID procedure. Screen pairs classified in the {\sc{High}} group exhibited a high amount of pixel difference according to PID and thus a larger number of GUI-changes, whereas the {\sc{Low}} group exhibited a low amount of pixel difference according to PID and thus had a low number of GUI changes. We randomly sampled an even number from each group to provide for a varied set of GUI-changes.

\subsubsection{Metric Collection Procedure and GUI Change Ambiguities}

\begin{figure}
\centering
\includegraphics[width=0.85\columnwidth]{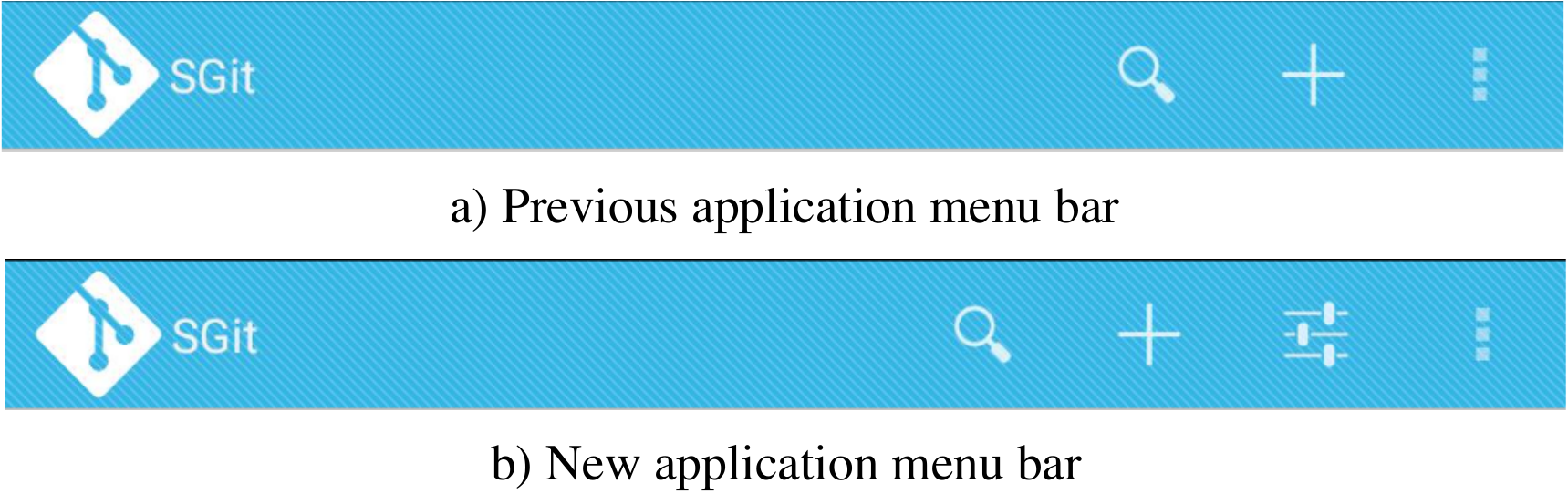}
\vspace{-0.4cm}
\centering
\caption{Illustration of a Potential Ambiguous GUI-change}
\label{fig:ambiguity}
\end{figure}

	In order to collect these metrics, it is necessary to manually examine each pair of corresponding screens between versions and generated reports. However, this is a very expensive manual procedure that involves evaluators visually examining screenshots, and inspecting GUI-metadata in order to calculate the metrics listed above. Furthermore, classifications of GUI-changes between subsequent versions of an application are open to multiple subjective interpretations, which may impact their calculation. For example, take the menu bar of the Sgit application shown in Figure \ref{fig:ambiguity}. This GUI-change is relatively simple, an additional icon \img{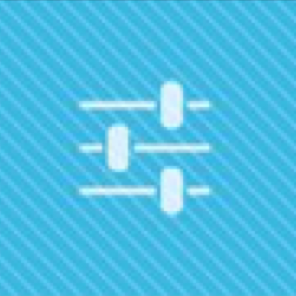} was added in the new app version. However, this could be interpreted in multiple manners. For example, one interpretation may be that the \img{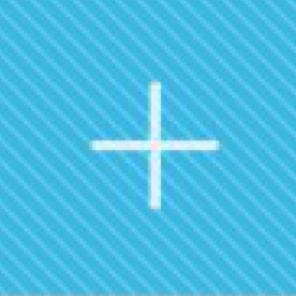} icon was changed to the \img{img/settings.png} icon, a new \img{img/plus.png} was added, and the \img{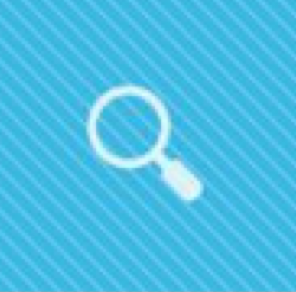} icon was shifted to the left. Another interpretation may be that the \img{img/settings.png} icon was added and the \img{img/plus.png} and \img{img/magnify.png} icons were shifted to the left.  Both of these are valid interpretations of the GUI change.  In fact, during our experimental investigation we came across different types of \textit{GUI Change Ambiguities} that may lead to multiple interpretations.  We forgo a detailed discussion of all ambiguities, but provide descriptions and examples in our online appendix~\cite{online-appendix}.

	In order to effectively collect the evaluation metrics listed earlier, we took several measures to ensure accurate calculations. First, we developed a tool that parses information from both screens in a corresponding screen pair and prints formatted readouts that list each GUI-component, its corresponding spatial metadata, a cropped image of the component, and the PID output in order to help in manual calculation of these metrics. Second, for RQ$_2$ and RQ$_3$, we simply accepted \textit{any} correct interpretation of a GUI change as correct when calculating the $DP$ and $CP$. Third, we employed multiple independent manual evaluators to examine the reports and screen pairs to calculate the metrics. More specifically, two authors independently calculated the metrics for \textit{each} screen pair. Then a third author evaluated the responses from the first two evaluators, and determined the final calculations from the combined responses.

\vspace{-0.1cm}
\subsection{RQ$_3$: Measuring Developer Performance in Detecting and Classifying GUI Changes}
\label{subsec:rq3-methodology}

	The main \textit{goal} of this RQ is to compare the effectiveness of \Gcat to developers at detecting and classifying GUI changes. In order to carry out this comparison, we needed to examine how well developers are able to comprehend and report changes in the GUIs of mobile apps. Thus, we conducted a user study in the form of an online survey consisting of four major components: (i) A \textit{Background} component that introduced the concept of a GUI change and our taxonomy of GUI-change types; (ii) A \textit{Demographic} component that asked participants about their programming background; (iii) A \textit{GUI-comprehension} component that asked participants to examine pairs of screenshots containing changes and document these changes; and (iv) a \textit{Report Feedback} component that asked participants to examine \Gcat reports and answer questions about their usefulness. The \textit{GUI-comprehension} component of this user study helps to answer RQ$_3$, whereas the \textit{Report Feedback} component helps to answer RQ$_4$ and is explained in further detail in the next subsection.  20 faculty and graduate student participants with at least three years of programming knowledge were recruited across 3 different universities.

	To derive the screen pairs to be used in both parts of the user study, three authors executed concurrent corresponding versions of each of the 31 apps and identified at least one screen pair between the two versions that contained a GUI change. For each identified screen pair with GUI-changes, a screenshot and GUI-metadata file were extracted using the Android Debugging Bridge's (\texttt{\small adb}) \texttt{\small screencap} utility and the \texttt{\small uiautomator} framework respectively.  This resulted in a set of 50 app screen pairs. Note that we did not use \CrashScope and \Gcats filtering procedure to produce this set of screens, in order to control the quantity of pairs for the user study.

	Given that the set of screen pairs extracted for the user study were taken from subsequent versions of real apps, the extent to which the GUI changed varies across the dataset.  However, in the context of the \textit{GUI-comprehension} user study, we want to understand the extent to which each participant can comprehend both simple and complex GUI-changes. Thus, similar to the procedure used in RQ$_2$, we divded the screen pairs into three groups according to the PID score. For the \textit{GUI-comprehension} component of the user study, we randomly selected 30 screen pairs from the candidate set of 50, ensuring that the 30 screens were evenly distributed across the three GUI-change groups.  Each participant in the study was assigned 3 screens from this set and the screens were assigned in such a manner that each screen was evaluated by two participants, each participant evaluated one {\sc{High}}, one {\sc{Medium}}, and one {\sc{Low}} from each GUI-change group, and the order in which the screens were presented to participants was randomized.

	During the \textit{GUI-comprehension} component of the survey, participants were asked to examine each screen pair and report each GUI-change according to the taxonomy presented at the beginning of the survey.  The GUI change taxonomy to be used was repeated on the survey screen where participants described the GUI-changes for reference. Each screen pair was accompanied by a text input box where participants were instructed to record one GUI-change per line in the form, $<$\textit{GUI-Change category}$>$:$<$\textit{Description of the GUI change}$>$. After all survey responses were collected, the $DP$, $CP$, and $R$ for each participant was calculated.  Three authors derived the ground truth and the evaluation metrics for the set of user study screens following the same methodology as in Sec. \ref{subsec:rq2-methodology}.

\vspace{-0.2cm}
\subsection{RQ$_4$: Investigating Perceived Developer Usefulness of \Gcat Reports}
\label{subsec:rq4-methodology}

	The \textit{goal} of this RQ is measure the perceived developer utility of \Gcat reports. This was carried out through the \textit{Report Feedback} component of the user study survey.  For this component of the survey, each participant was shown two screen pairs, and the corresponding \Gcat report for these screens. The participants were then asked 5 Likert-based \textit{user experience (UX)} and five free-response \textit{user preference (UP)} questions, which were derived from SUS usability scale introduced by Brooke~\cite{Brooke:96}, and the user experience honeycomb by Morville~\cite{Morville:04} respectively. The screen pairs for the \textit{Report Feedback} component of the user study survey were comprised of the 20 remaining screens after the sampling for the \textit{GUI-Comprehension} component.  Screens were assigned to participants in such a manner that each screen pair and report were evaluated by two participants, screen pairs were distributed as evenly and randomly as possible across the GUI-change groups.

\vspace{-0.2cm}
\section{Empirical Results}
\label{sec:results}

\subsection{RQ$_1$: Performance of Screen Filtering and Matching}
\label{subsec:results-rq1}

	Our first RQ investigates the performance of \Gcats screen filtering and matching procedure. Running \CrashScope through all 61 of our subject \texttt{\small apks} resulted in 3,854 total extracted screens, or $\approx$ 63 screens per \texttt{\small apk}.  \Gcats filtering procedure was able to reduce this set to a much more manageable 316 screens for the matching procedure. This results in an $FS$ measurement of $(3,854-316/3,2854)*100=91.8\%$, meaning that over 90\% of the collected screens were filtered out as redundant, drastically reducing the information burden on developers for reading GUI-change reports. These filtered screens resulted in 158 matched screen pairs, which exhibited a \textit{Matching Precision (MP)} of 84.8\%. This illustrates that \Gcat is able to both effectively filter and match corresponding screen pairs that were automatically extracted from automated dynamic analysis of subsequent app versions. 

\vspace{-0.2cm}
\subsection{RQ$_2$: \Gcat Performance}
\label{subsec:results-rq1}

\begin{figure}
\centering
\includegraphics[width=\columnwidth]{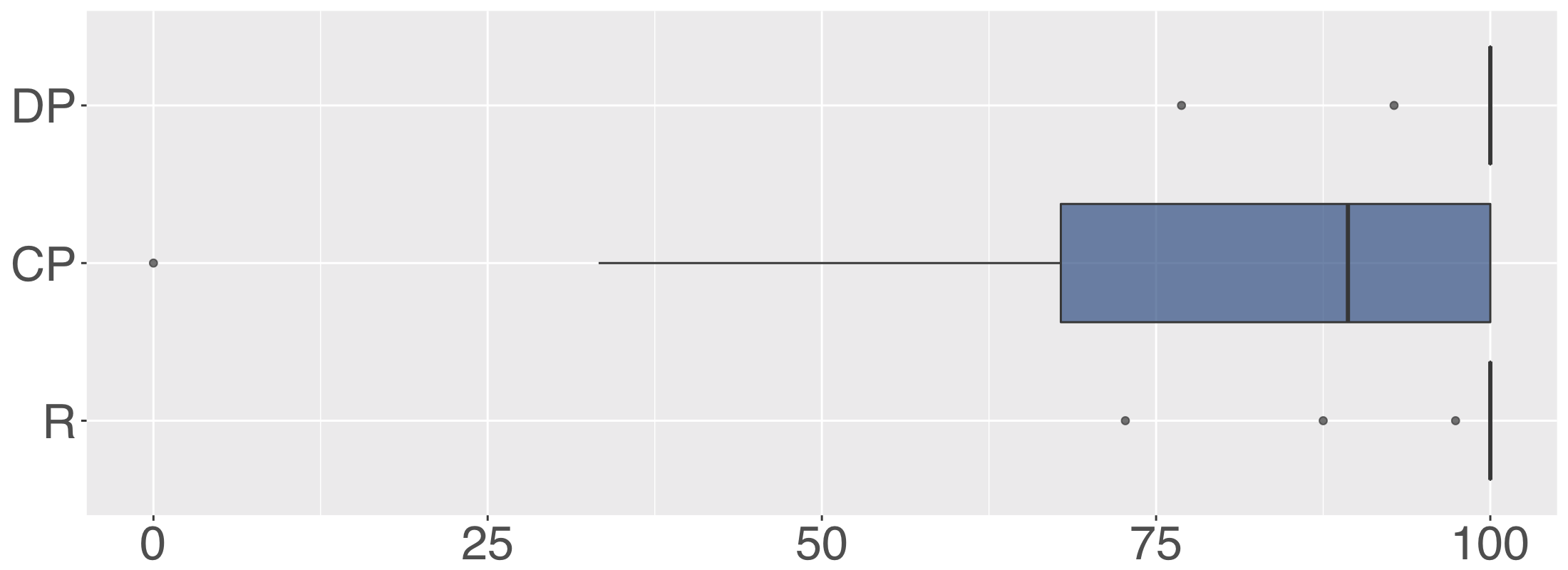}
\vspace{-0.7cm}
\centering
\caption{\Gcat $DP$, $CP$, and $R$}
\label{fig:gcat-performance}
\end{figure}

	Running \Gcats change analysis over the 158 matched screen pairs took an average of 13.1 seconds per screen pair. The \Gcat results for $DP$, $CP$ and $R$ are illustrated as box-plots across the analyzed reports in Figure \ref{fig:gcat-performance}. \Gcat is able to achieve an average $DP$ of 98.3\% and an average $R$ of 97.6\%, however $CP$ is a bit lower than these with an average value of 76.7\%. This means that \Gcat is able to effectively detect GUI-changes with few false positives, and rarely misses reporting existing GUI-changes on a screen. However, when classifying these GUI-changes into their corresponding taxonomy categories, there were certain cases of incorrect classification.

	The largest source of false positives in terms of the $CP$ came from ambiguities related to Font changes and Font Color changes. As explained in Section \ref{subsec:gui-change-detection}, \Gcat derives a color histogram (CH) from cropped images of textual components, and if the Euclidean distance between these Color Histograms does not match within a given threshold, then a Font color change is reported. However, the sensitivity of this threshold can vary between different styles of fonts, making it difficult to properly tune. This results in several Font color violations being classified as Font style changes.  However, it should be noted that these classifications are very similar and are less impactful to the utility of reports than if a more orthogonal classification was made (\eg Font color $\rightarrow$ Layout Change).

\vspace{-0.2cm}
\subsection{RQ$_3$: Developer Performance}
\label{subsec:results-rq1}

\begin{figure}
\centering
\includegraphics[width=\columnwidth]{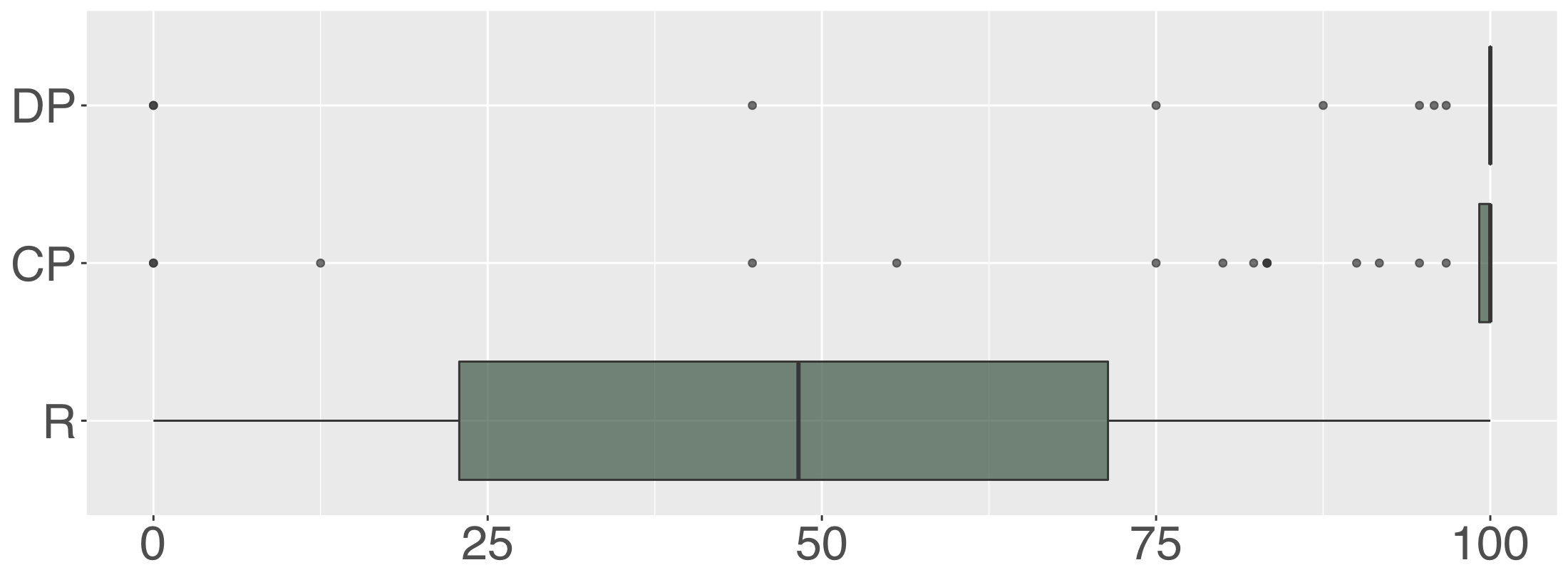}
\vspace{-0.7cm}
\centering
\caption{Developer $DP$, $CP$, and $R$}
\label{fig:developer-performace}
\vspace{-0.3cm}
\end{figure}

	The developer results for $DP$, $CP$ and $R$ are illustrated as a box-plots across the analyzed reports in Figure \ref{fig:developer-performace}. On average, developers achieved a $DP$ of 94.9\%, and a $CP$ of 91.72\%. However, their recall suffered quite a bit, with developers on average only reporting 49.4\% of existing GUI-changes for a given screen pair. Furthermore, on average developers required 9 minutes and 8 seconds to detect and classify the GUI changes for the three assigned screen pairs. In general this means that, while developers were generally accurate at reporting and classifying changes when they recognized them, there were a large number of changes that were not reported, and the reporting process was time consuming. The underlying reason for missed changes varied across developers and screen pairs. In certain cases, subtle changes in the layout or size of components were not reported, however, in other cases, more easily observable changes were missed, including the failure to report entirely new or removed components between screen pairs. When comparing the developer's performance to \Gcat, we find that \Gcat outperformed developers in each metric.

\subsection{RQ$_4$: Perceived Utility of \Gcat reports}
\label{subsec:results-rq1}

\begin{figure}
\centering
\includegraphics[width=\columnwidth]{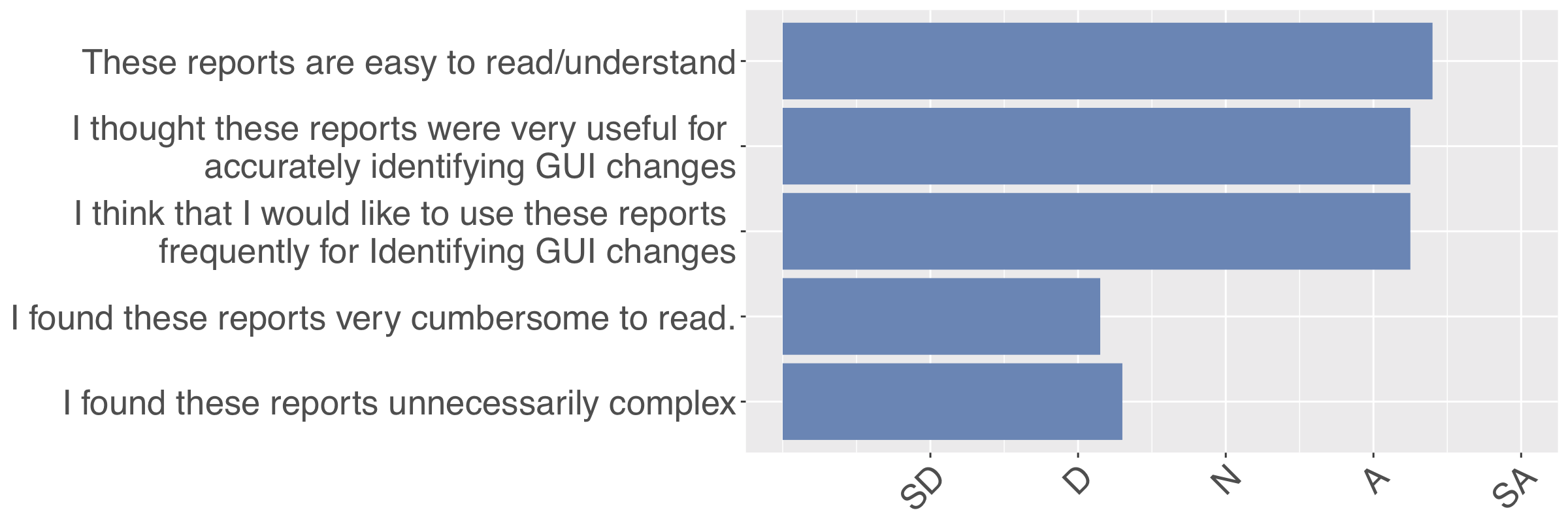}
\vspace{-0.7cm}
\centering
\caption{Average Developer UX Question Responses. SD=Strongly Disagree, D=Disagree, N=Neutral, A=Agree, SA=Strongly Agree}
\label{fig:developer-utility}
\end{figure}

The results for the UX questions used to measure the developer perceived usefulness of the \Gcat reports are given as average values in Figure \ref{fig:developer-utility}. These results are generally very positive, with developers agreeing on average that the \Gcat reports are (i) easy to understand, (ii) useful for identifying GUI changes, and (iii) applicable for frequent use. They also generally found that \Gcat reports were \textit{not} cumbersome to read or overly complex.  These responses help to illustrate the utility that developers found when examining reports. 

	For the User Experience Questions ($UX$), we asked participants about four aspects of the reports: (i) the information that was most useful, (ii) what additional information would have been helpful, (iii) the elements they liked the most from the reports, and (iv) the elements they liked the least from the reports. In response to the first question, many users indicated that they found the Full Annotated Screenshots, and detailed list of GUI-component changes the most useful. For example one participant responded, \textit{``The detailed changes, since they show the status of the elements before/after the changes.''}, whereas another participant indicated, \textit{``Seeing how components moved with the highlighted red boxes.''}. The developers also gave some useful feedback for improvements to the tool. For example, \textit{``A way to group the changes or a potential importance (e.g., a new component may be more important as compared to a 2px change).''}. In responding to which elements they liked the most, the responses mostly echoed the first question, where the side-by-side annotated screenshots and detailed list of GUI changes were the most cited. Finally, while some participants indicated that they did not dislike any of the information in the reports, others cited the NL summary, and tree comparison as areas for improvement. We provide the full set of responses in our online appendix~\cite{online-appendix}.

\vspace{-0.2cm}
\section{Limitations \& Threats to Validity}
\label{sec:threats}

\noindent \textbf{Limitations:} Our experimental evaluation of \Gcat has shown the tool achieves remarkable effectiveness, however, the approach does exhibit certain limitations that serve as motivation for future work. Currently our approach may not properly handle dynamic screen content. For example, a list that is loaded over the network that might not \textit{actually} change between versions could be detected as a series of GUI-changes due to differing content. This problem could be mitigated by asking developers to annotate certain screen content as dynamic, or through automatic recognition of dynamic content via machine learning. Second, currently our approach operates only on native Android apps, and has not been implemented or tested for iOS or hybrid apps. However, we expect the underlying techniques for detecting and classifying GUI changes to apply to other types of apps and platforms, where the largest challenges lie in engineering methods to extract accurate GUI metadata. Finally, our study of \Gcats screen matching algorithm revealed limitations of our approach, as $\approx 15\%$ of the screens were not correctly matched. Future work could look towards exploring more sophisticated matching algorithms that take greater advantage of certain structural properties of GUI-metadata.

\noindent \textit{\textbf{Internal Validity: }} In our experiments evaluating the \Gcat approach, internal threats may arise from our manual examination of reports (RQ$_2$), and responses from users (RQ$_3$).  However, three authors independently examined all reports and user responses following a set, rigorous methodology. Also, our results illustrate clear trends that we expect would hold across different evaluators. 

\noindent \textit{\textbf{Construct Validity:}} One threat to construct validity concerns differences in the sets of screen pairs utilized to investigate RQ$_2$ and RQ$_3$. In answering RQ$_2$ we used randomly sampled screen pairs that were automatically derived from \Gcats automated GUI exploration engine. This study was carried out in this way to evaluate \Gcat in its intended, fully automated use case.  However, for the user study, we needed more control over the number of screen pairs in order to design the screen pair assignment for participants, and thus we manually extracted screens from our set of subject applications that had known differences. However, the sampling procedure based on PID described in Section \ref{sec:study-design} ensured that a similarly varied set of screens was used between the two studies, mitigating this threat to validity concerning our experimental observations.

\noindent \textit{\textbf{External Validity:}} We utilized a set of 31 open source subject applications from the F-Droid marketplace in our experimental evaluation of \Gcat. There is the potential that the experimental results observed in this paper may not generalize to a larger set of applications, or that the GUIs of the open source applications studied differ from those of paid apps on Google Play. However our set of subject applications represent varying sizes and popularities of apps. Thus we assert that our subject set of applications is varied enough to draw meaningful experimental conclusions.  Another threat to external validity concerns the generalization of the results of our developer survey to a broader set of mobile developers. While our participants primarily came from academic backgrounds, they had an average general programming experience of 6.8 years and an average mobile programming experience of 1.5 years. Furthermore past work has found responses from such studies to be representative of professional developers~\cite{Salman:ICSE'15}.

\vspace{-0.2cm}
\section{Related Work}
\label{sec:back-related}

There is a sizable body of existing that aims to automatically summarize code-related information, such as methods and release notes~\cite{Robillard:ICSME'17,Moreno:ICSE'15,Moreno:FSE'14,Moreno:ICPC'13a,Moreno:ICPC'13,Moreno:TSE'17}. However, we forgo a detailed discussion of these techniques as they do not specifically attempt to summarize aspects of GUIs.

\noindent \textit{\textbf{GUI Differencing:}} The most closely related work to ours is that by Xie \etal who introduced a GUI differencing approach called {\sc Guide}~\cite{Xie:ICSM'09}.  {\sc Guide} is capable of resolving mappings between GUI objects of GUI hierarchy trees in different app versions, however, its matching procedure is not described in detail.  While {\sc Guide} is capable of deriving GUI mappings, it is not capable of detecting, reporting or summarizing GUI-changes that occur between these mappings. Furthermore, the effectiveness of {\sc Guide} was not evaluated on a large dataset of apps with existing GUI-changes. 

\noindent \textit{\textbf{Detecting Presentation Failures in Mobile \& Web Apps:}} A growing body of work has been dedicated to detecting \textit{presentation failures} and \textit{}design violations in mobile and web apps.  Moran \etal introduced \GVT~\cite{Moran:ICSE'18}, which is capable of detecting design violations and presentation failures that occur between a mock-up of an app's GUI and its implementation of that mock-up.  While this approach shares similarities with \Gcat, there are several key differences.  First, rather than resolving information between a GUI mock-up and an implementation of that mock-up, \Gcat must resolve information between subsequent app versions. Second, whereas \GVT requires the manual specification of screens to compare, \Gcat  derives these screens automatically via automated GUI exploration of an app. Third, \Gcat aims to support comprehension tasks, and thus must effectively summarize the GUI changes both visually and in natural language. There is also an existing body of work that aims to detect, classify, and fix presentation failures in web apps~\cite{Mahajan:ICST'15,Mahajan:ICST'16,RoyChoudhary:ICSE'13,Mahajan:ISSTA'17}.  However, these approaches do not target mobile apps, and are not concerned with summarizing GUI changes in evolving apps.

\noindent \textit{\textbf{Cross-Browser Testing:}} There also exist approaches for XBT, also known as cross browser testing \cite{RoyChoudhary:ICSE'13,Choudhary:ICST'12,RoyChoudhary:ICSM'10}, that are capable of detecting and reporting differences between web pages rendered in different types of browsers. While this work shares some underlying goals with our approach (\eg detecting corresponding screens, GUI elements), \Gcat exhibits a few notable departures that illustrate its novelty. First, in order to effectively summarize evolutionary GUI changes \Gcat is capable of \textit{classifying} detected changes into common change categories for mobile app GUIs. Second, our approach is able to generate human-readable reports that contain natural language summary changes at multiple granularities.

\vspace{-0.1cm}
\section{Conclusion \& Future Work}
\label{sec:conclusion}

	We present \Gcat, an automatic summarization tool used for detecting and reporting GUI changes during the evolutionary development of mobile apps. An evaluation of \Gcat illustrates that our approach is effective, outperforming developers, and reports useful information in a comprehensible manner. Our future work entails a more precise classification for GUI changes as well as continuing to improve the quality of the NL summarizations. Additionally, we aim to enable \Gcat to effectively analyze and classify dynamic screen content.

\begin{acks}
This work is supported in part by the NSF CCF-1815186 grant. Any opinions, findings, and conclusions expressed herein are the authors' and do not necessarily reflect those of the sponsors. The authors would like to thank the ASE'18 reviewers whose insightful comments which greatly improved this paper.
\end{acks}

\balance
\bibliography{ms}
\bibliographystyle{abbrv}

\end{document}